\documentclass[usenatbib,usegraphicx]{mn2e}
\usepackage{amssymb}
\usepackage[fleqn]{amsmath}
\usepackage{charter}
\usepackage{mathdesign}
\usepackage{url}
\usepackage{bookmark}
\providecommand{\eprint}[1]{\href{http://arxiv.org/abs/#1}{#1}}
\providecommand{\adsurl}[1]{\href{#1}{ADS}}
 
\usepackage{natbib}

\newcommand{\na} {New Astronomy}

\newcommand{\pre} {Phys. Rev. E}
\newcommand{\memsai} {Mem.~Soc.~Astron.~Italiana}

\usepackage{graphicx}
\usepackage{subfigure}

\usepackage{color}
\usepackage[usenames,dvipsnames]{xcolor}

\graphicspath{{figures/}}

\pagerange{\pageref{firstpage}--\pageref{lastpage}} \pubyear{2011}

\def\LaTeX{L\kern-.36em\raise.3ex\hbox{a}\kern-.15em
T\kern-.1667em\lower.7ex\hbox{E}\kern-.125emX}

\newcommand{\B}{\begin{eqnarray}}
\newcommand{\E}{\end{eqnarray}}

\newcommand{\bm}[1]{\mbox{\boldmath{$#1$}}}
\newcommand{\del}{{\bf \nabla}}

\title[Magnetic Recollimation in Relativistic Jets]{Magnetic Domination of Recollimation Boundary Layers in Relativistic Jets}

\author[Kohler \& Begelman]
{Susanna Kohler$^{1, 2 \ast}$ and
Mitchell C. Begelman$^{1, 2 \star}$
\\$^1$ JILA, University of Colorado and National Institute of Standards and Technology, Boulder, CO 80309-0440, USA 
\\$^2$ Department of Astrophysical and Planetary Sciences, University of Colorado, Boulder, CO 80309-0391, USA
\\Email: $^\ast$ kohlers@colorado.edu,
$^\star$ mitch@jila.colorado.edu
}

\begin{document}

\label{firstpage}

\maketitle

\begin{abstract} 

We study the collimation of relativistic magnetohydrodynamic jets by the pressure of an ambient medium, in the limit where the jet interior loses causal contact with its surroundings. This follows up a hydrodynamic study in a previous paper, adding the effects of a toroidal magnetic field threading the jet. As the ultrarelativistic jet encounters an ambient medium with a pressure profile with a radial scaling of $p~\propto~r^{-\eta}$ where $2<\eta<4$, it loses causal contact with its surroundings and forms a boundary layer with a large pressure gradient. By constructing self-similar solutions to the fluid equations within this boundary layer, we examine the structure of this layer as a function of the external pressure profile. We show that the boundary layer always becomes magnetically dominated far from the source, and that in the magnetic limit, physical self-similar solutions are admitted in which the total pressure within the layer decreases linearly with distance from the contact discontinuity inward. These solutions suggest a `hollow cone' behavior of the jet, with the boundary layer thickness prescribed by the value of $\eta$. In contrast to the hydrodynamical case, however, the boundary layer contains an asymptotically vanishing fraction of the jet energy flux.

\end{abstract}


%%%%%%%%%%%%%%%%%%%%%%%%%%%%
%%%%%%%%%%%%%%%%%%%%%%%%%%%%
%%%%%%%%%%%%%%%%%%%%%%%%%%%%

\section{Introduction}

The outflows from active galactic nuclei (AGN) are thought to be highly relativistic (\citealp{MB84}) and highly collimated (e.g. \citealp{Jorstad05}), but the cause of this collimation is uncertain.

Because jet-launching is generally believed to be electromagnetically driven (e.g. \citealp{Blandford82,Contopoulos94}), one of the most commonly-accepted explanations for the observed collimation is that jets are threaded with magnetic fields that cause collimation via magnetic tension (e.g. \citealp{Benford78,MB95}). Supporting this theory, it has been demonstrated that both relativistic and non-relativistic hydromagnetic outflows must eventually become collimated (\citealp{Chiueh91,Heyvaerts89}). For magnetic fields acting alone, however, collimation will only happen on extremely large scales (\citealp{Eichler93,MB94,MB95}). 

To cause jets to collimate on reasonable scales, there must be an additional mechanism at work. A logical culprit is confinement by the pressure of an external medium. Pressure confinement has been demonstrated to act effectively on its own (e.g. \citealp{Levinson00,BL07,Kohler12}), and accretion disk winds surrounding an AGN provide an ideal ambient medium to help to collimate the jet.

There have been many numerical studies of magnetized jets (e.g. \citealp{KomissarovNumSim99,Hawley06,Beckwith08a,McKinney09}), with the goal of forming a self-consistent description of the jet-launching and collimation mechanisms. These numerical simulations have several restrictions, however, one of which being that the boundary of the jet, rather than having its shape determined by pressure balance, is generally treated as a rigid wall (e.g. \citealp{KomBarkVla07,KomVlaKon09,Komissarov11,Tchekhovskoy10}). This construction doesn't allow the ambient pressure to affect collimation of the jet.

Treatments that do include effects of the external medium commonly focus on describing jets that remain in causal contact (e.g. \citealp{Zakamska08,Lyubarsky11}). As an ultrarelativistic jet expands into an ambient medium with a pressure profile $p \propto r^{-\eta}$, it will eventually lose causal contact if $\eta > 2$ and the opening angle is greater than $1/\Gamma$, where $\Gamma$ is the bulk Lorentz factor of the fluid. Observations of gamma-ray bursts indicate that these relativistic jets largely have opening angles greater than $1 / \Gamma$ (e.g. \citealp{Piran04}; see \citealp{Tchekhovskoy10} for discussion), and AGN outflows with large Lorentz factors may similarly be causally disconnected; thus the poorly-studied regime of a jet that has lost causal contact is of physical interest.

In a previous paper (\citealp{Kohler12}, hereafter KBB12) we developed a model describing the recollimation boundary layer of a purely hydrodynamic, ``hot'' (pressure-dominated) jet with an ultrarelativistic equation of state. In this model, we assumed that the pressure outside the jet decreases with $r$ so rapidly that the jet interior loses causal contact with its boundary, resulting in a shocked boundary layer forming within the jet. Though the jet interior is causally disconnected, the boundary layer is nevertheless narrow enough to remain in causal contact itself. Assuming self-similarity as a function of $r$, we calculated how the transverse structure of the jet boundary layer depends on the value of $\eta$ in the external pressure profile.

We now expand this work to include, in addition to collimation by the external medium, the effects of a magnetic field within the jet. We include only a toroidal field, as it is the toroidal field that dominates the dynamics at large radii, far outside the light cylinder (\citealp{MB84,Contopoulos95,Beskin09}). In Section 2, we first demonstrate that seeding a jet with a magnetic field at the base will always cause it to become magnetically dominated at large radii. We then find self-similar solutions for the boundary layer of the jet in the limit of magnetic dominance. In Section 3 we discuss the results, and in Section 4 we conclude. 

%%%%%%%%%%%%%%%%%%%%%%%%%%%%%%
%%%%%%%%%%%%%%%%%%%%%%%%%%%%%%

\section{Self-Similar Treatment of the Magnetized Boundary Layer}

We use spherical coordinates to model a hot, ultrarelativistic jet that is symmetric about the $z$-axis and has approximately radial streamlines -- an approximation justified because the jet interior is causally disconnected from the environment. We assume that the jet is injected from a point source with steady flow, and we examine the jet in its steady-state configuration.

We focus on modeling the boundary layer of jet material that forms at the interface between the jet and the stationary ambient medium. This layer is bounded on the inside by a shock front or a rarefaction front, and on the outside by a contact discontinuity. There is no mass flux across the contact discontinuity, and the pressure must be matched on either side of it. The physics of the ambient medium is wrapped into the external pressure profile $p_e$.

We adopt a pressure profile for the ambient medium of $p_e \propto r^{-\eta}$, fixing the pressure external to the jet to be dependent only on the parameter $\eta$. We focus on the case of ambient pressure where $2 < \eta < 4$, as in our work in KBB12 or in, e.g., \citet{BL07}, because the jet interior is out of causal contact with the exterior for this range in $\eta$. Physically, this pressure profile range could describe a confining wind, an accretion flow, a disk corona, or even a stellar envelope in a GRB collapsar model (e.g. \citealp{Bromberg11}).

As in KBB12, we assume that the opening angle of the jet is much greater than $1/\Gamma$, such that causal contact has been lost. We construct a boundary layer that remains in causal contact, such that its thickness is of order $\Delta \theta \sim 1/\Gamma$. The boundary layer is thus very thin compared to the width of the jet. The radius of curvature of the jet is then much larger than the width of the boundary layer, allowing us to treat the curvature as a small effect. 

We begin with the equations for relativistic MHD (RMHD) in flat spacetime (e.g. \citealp{Dixon78,KomissarovNumSim99,Zakamska08}), including the effects of a toroidal magnetic field within the jet. We ignore rotation since our regime of interest is far outside the light cylinder and, indeed, far outside the fast magnetosonic surface, rendering rotation effects unimportant. We again assume an ultrarelativistic equation of state such that the total proper energy density is given by $\epsilon = \rho + 3p \approx 3p$, where $\rho$ and $p$ are, respectively, the proper rest mass density and the pressure of the fluid within the boundary layer.

The continuity equation remains unchanged with the addition of a magnetic field, 
	\begin{align}
	\del \cdot (\rho \bm \beta \Gamma) &= 0,  \label{continuity}
	\end{align}		
where $\beta$ and $\Gamma$ are the velocity ($\bm{\beta} = \bm{v}/c$) and the bulk Lorentz factor of the fluid within the boundary layer. Assuming an energy equation of $p \propto \rho^{4/3}$ and taking $\theta \sim$ constant, the continuity equation becomes
	\begin{align}
	\frac{1}{r^2} \frac{\partial}{\partial r} ( r^2 p^{3/4} \Gamma \beta_r) + \frac{1}{r} \frac{\partial}{\partial \theta} (p^{3/4} \Gamma \beta_\theta) &= 0
	\end{align}
in spherical coordinates.
	
Denoting the observer-frame magnetic field within the jet boundary layer as $\bm{B}$, and using the ideal MHD condition to express the observer-frame electric field as $\bm E = -\bm v \times \bm B$, the momentum equation can now be written as
	\begin{align}
	(4 p \Gamma^2 + B^2) (\bm \beta \cdot \del) \bm \beta + \del (p + \frac{1}{2} B^2 \Gamma^{-2} ) \nonumber\\
	 -\bm B [\del \cdot (B \Gamma^{-2})] - \Gamma^{-2} (\bm B \cdot \del ) \bm B &= 0,    \label{mom}
	\end{align}
where the parallel and perpendicular components of the momentum equation are obtained by taking the respective vector dot and cross product of $\bm \beta$ with Eq \eqref{mom}.	

Finally, we add the equation for flux freezing,
	\begin{align}
	\del \times (\bm \beta \times \bm B) &= 0.    \label{fluxfreeze}
	\end{align}
Writing this in spherical coordinates, and assuming a toroidal magnetic field $\bm B = B(r, \theta) \bm {\hat{\phi}}$, we have
	\begin{align}
	\frac{\partial}{\partial r} (r \beta_r B) + \frac{\partial}{\partial \theta}(\beta_\theta B) &= 0.
	\end{align}
	
The full form of these four equations admit self-similar solutions only in the case where $\eta =4$, which corresponds to a quasi-monopole flow with no collimation and therefore no distinct boundary layer. For all other values of $\eta$, the full equations allow only for trivial solutions due to overconstraint of the system. We now demonstrate, however, that it is not appropriate to use these equations in their full form. They should instead be examined in the magnetically-dominated limit -- where they do admit non-trivial self-similar solutions.

%%%%%%%%%%%%%%%%%%%%%%%%%%%%%%

\subsection{Demonstration of Asymptotic Magnetic Dominance}

Suppose that the magnetic field takes the form $B~=~r p^{3/4} \Gamma f(r, \theta)$ where $f$ is some function of $r$ and $\theta$. Inserting this into Eq \eqref{fluxfreeze}, we obtain
	\begin{align}
	\del \cdot (p^{3/4} \Gamma \bm \beta f) &= 0
	\end{align}
which can be expanded and combined with Eq \eqref{continuity} to show that
	\begin{align}
	\bm \beta \cdot \del f &= 0,
	\end{align}
indicating that the function $f$ must be constant along streamlines.

Because the contact discontinuity is a streamline, we can therefore state that $B \propto r p^{3/4} \Gamma$ along the contact discontinuity. As in \citet{Zakamska08}, we now define a magnetization parameter $\beta_B$ as the ratio of magnetic to gas pressure (the inverse of the usual plasma beta). Along the contact discontinuity this parameter is thus given by
	\begin{align}
	\beta_B = \frac{B^2}{p \Gamma^2} \propto r^2 p^{1/2}. \label{sigma}
	\end{align}

Because pressure must be matched across the contact discontinuity, the external pressure $p_e \propto r^{-\eta}$ must be balanced at that point by the total internal pressure within the boundary layer, $p_{tot} = p + B^2/\Gamma^2$. We now examine $\beta_B$ in two extreme cases: the limit where the pressure balance at the contact discontinuity is supplied solely by the gas pressure within the boundary layer, and the limit where the balance is supplied solely by the magnetic pressure within the layer.

In the gas pressure-dominated case, the internal gas pressure $p$ is equivalent to the external pressure $p_e$, and must therefore scale in the same way, such that $p \propto r^{-\eta}$. 
Applying this to Eq \eqref{sigma} demonstrates that in this case, $\beta_B \propto r^{(4 - \eta)/2}.$ Thus, for $2 < \eta < 4$, $\beta_B$ scales as $r$ to some positive power.

In the magnetically-dominated case, the external pressure is balanced by the magnetic pressure such that $B^2 / \Gamma^2 \propto r^{-\eta}$. This scaling implies that the internal gas pressure is given by $p \propto r^{-(2/3)(\eta+ 2)}$, and we obtain $\beta_B \propto r^{(4-\eta)/3}$. Again, for $2 < \eta < 4$, $\beta_B$ scales as $r$ to some positive power.

Thus we see that in both extreme cases, $\beta_B$ grows with increasing $r$. This suggests that no matter how small a magnetic field the jet is seeded with, the boundary layer will eventually become magnetically-dominated far from the jet source. With this in mind, we now repeat the calculations performed in KBB12 with the inclusion of a toroidal magnetic field, specifically in the limit where $\beta_B \gg 1$.

%%%%%%%%%%%%%%%%%%%%%%%%%%%%%%

\subsection{Solutions in the Magnetic-Dominance Limit}

We first rederive the fluid equations in Section 2.1 in the limit where $\beta_B \gg 1$. Continuity and flux freezing are unchanged, but terms in the momentum equation containing $1/\beta_B$ are negligible in this limit.

We now make scaling arguments as in KBB12: we assume $\beta_\theta$ is of order $1/ \Gamma$, since this is the maximum transverse speed that can be achieved without a shock forming, and $\beta_r$ is of order one. With this characteristic scale, $\frac{\partial}{\partial \theta} \sim \Gamma \frac{\partial}{\partial r} $. Expressing $\beta_r$ in terms of $\beta_{\theta}$ and $\Gamma$ and employing the fact that $\beta_{\theta}^2 + \Gamma^{-2} \ll 1$, we have $\beta_r \approx 1- \frac{1}{2} ( \beta_{\theta}^2 + \Gamma^{-2})$. Using these scalings and keeping terms only to lowest order, the parallel and perpendicular components of the momentum equation are:
	\begin{align}
	r \dfrac{\partial p}{\partial r} + \beta_\theta \dfrac{\partial p}{\partial \theta} + \dfrac{B}{\Gamma^2} \left( B + r \dfrac{\partial B}{\partial r} + \beta_\theta \dfrac{\partial B}{\partial \theta} \right) = 0 \\
	B^2 \left( r \dfrac{\partial \beta_\theta}{\partial r}  + \beta_\theta + \beta_\theta \dfrac{\partial \beta_\theta}{\partial \theta} - \dfrac{1}{\Gamma^{3}} \dfrac{\partial \Gamma}{\partial \theta} \right) + \dfrac{\partial p}{\partial \theta} + \dfrac{B}{\Gamma^{2}} \dfrac{\partial B}{\partial \theta} = 0.
	\end{align}
We now attempt to construct self-similar solutions in the following fashion:
	\begin{align}
	\dfrac{1}{\Gamma} &= g(\xi) r^{-x},&  \quad \beta_\theta &= h(\xi) r^{-x}, \nonumber \\
	B &= b(\xi) r^{x - \eta/2},&  \quad p &= a(\xi) r^{-\alpha} ,
	\end{align}
such that the external gas pressure is matched by the internal magnetic pressure, $B^2/ \Gamma^2 \propto r^{-\eta}$. In these solutions $x$ and $\alpha$ are constant free parameters describing the radial scaling, and $g$, $h$, $b$ and $a$ are functions of a similarity variable $\xi$ (as in KBB12) that describes the distance from the contact discontinuity, normalized by the expected scale of the boundary layer, $\xi \propto (\theta_c - \theta)/{\Delta \theta}$. The angular thickness of the boundary layer is expected to scale as $\Delta \theta = {1}/{\Gamma_c}$, such that $\xi \propto r^{x} (\theta_c - \theta)$, where $\theta_c = \theta_c(r)$ is the location of the contact discontinuity.

The fact that the streamlines at $\theta_c$ must be parallel to the contact discontinuity, requiring that $\beta_\theta(\theta_c) = r d\theta_c / dr$, yields the further constraint that
	\B
	\dfrac{d\theta_c}{dr} &=& h_0 r^{-(x+1)}, \label{dthetac}
	\E
where $h_0 = h(\xi = 0)$ is a negative constant for collimating solutions. While the flow is very nearly radial, this expression describes the small deviation of the flow lines resulting from subtle collimation. Choosing the proportionality constant such that $\xi$ is defined as
	\B
	\xi = - \dfrac{1}{h_0} r^{x} (\theta_c - \theta)
	\E
absorbs the boundary condition into the similarity variable and ensures collimating solutions (such that $h_0<0$). 

We now recast the fluid equations in terms of these functions. The continuity and flux-freezing equations are fully self-similar and become, respectively,
	\begin{align}
	\left( 3 \dfrac{a'}{a} - 4 \dfrac{g'}{g} \right) \left( x \xi - 1 + \dfrac{h}{h_0} \right) + \left( 4x + 8 - 3\alpha + 4 \dfrac{h'}{h_0} \right) &= 0 \label{ss1} \\
	\dfrac{b'}{b} \left( x \xi - 1 + \dfrac{h}{h_0} \right) + \left( x - \dfrac{\eta}{2} + 1 + \dfrac{h'}{h_0} \right) &= 0,   \label{ss4}
	\end{align}
where primes denote differentiation with respect to $\xi$.

Now examining the parallel component of the momentum equation, one can see that the radial scaling does not automatically vanish:
	\begin{align}
	\dfrac{b^2g^2}{a} \left[ 1 + x - \dfrac{\eta}{2} + \dfrac{b'}{b} \left(  x \xi - 1 + \dfrac{h}{h_0} \right) \right] \nonumber \\ 
	+ r^{\eta-\alpha} \left[ \dfrac{a'}{a} \left( x \xi - 1 + \dfrac{h}{h_0} \right) - \alpha \right] &= 0.
	\end{align}
Because we are specifically examining the regime where $\beta_B \propto B^2/(p\Gamma^2) \propto r^{\alpha - \eta} \gg 1$, however, we can assume that $r^{\eta-\alpha}$ will be very small for large $r$, rendering the second term negligible.

Using this logic, the parallel and perpendicular components of the momentum equation become, respectively,
	\begin{align}
	\dfrac{b^2g^2}{a} \left[ 1 + x - \dfrac{\eta}{2} + \dfrac{b'}{b} \left(  x \xi - 1 + \dfrac{h}{h_0} \right) \right] &= 0 \label{ss3} \\
	h(1-x) + h' \left(  x \xi - 1 + \dfrac{h}{h_0} \right) + \dfrac{g^2}{h_0} \left( \dfrac{g'}{g} + \dfrac{b'}{b} \right) &= 0. \label{ss2}
	\end{align}

Assuming that $b$, $g$, $a$ and $h$ are finite and non-zero, Eqs \eqref{ss4} and \eqref{ss3} imply that $h' = 0$. This assumption results in Eq \eqref{ss4} becoming
	\begin{align}
	\dfrac{b'}{b} &= - \dfrac{x - \eta/2 + 1}{x\xi} \label{ss4sing}
	\end{align}
and yields the following general solutions for the functions describing the transverse behavior within the boundary layer:
	\begin{align}
	h &= h_0 \\ 
	b &= A \xi^ {- (2+2x-\eta)/2x} \\
	g &= \pm \left[ B \xi^{(2+2x-\eta)/x} - \dfrac{ 2 h_0^2 x (x-1)}{2+x-\eta} \xi  \right]^{1/2} \\
	\dfrac{a'}{a} &= \dfrac{3\alpha - 4 - 2\eta}{3x} \xi^{-1} - \dfrac{4}{3} (1-x) \left(\dfrac{g}{h_0} \right)^{-2},
	\end{align}
where $A$ and $B$ are constants of integration that are defined by the boundary conditions.

To produce physical solutions, we examine the special case where we prevent $b'$ from having a singularity at $\xi = 0$ by setting $x-\eta/2 + 1=0$, implying that $b$ is a constant. In terms of boundary conditions $h_0, g_0, b_0,$ and $a_0$, which serve as scaling factors and allow us to determine our functions self-consistently, the physical solutions within the boundary layer are therefore
	\begin{align}
	h &= h_0 \\
	b &= b_0 \\
	g &= - h_0 \left[\left(\dfrac{g_0}{h_0}\right)^2 - (4-\eta)\xi \right]^{1/2} \\
	a &= a_0 \left( \dfrac{h_0}{g_0} \right)^{4/3} \left[\left(\dfrac{g_0}{h_0}\right)^2 - (4-\eta)\xi \right]^{2/3},
	\end{align}
with the constraints $x = \frac{\eta}{2} -1$ and $\alpha = \frac{2}{3} (\eta+2)$, such that
	\begin{align}
	\dfrac{1}{\Gamma} &= g r^{1-\eta/2},&  \quad \beta_\theta &= h r^{1-\eta/2}, \nonumber \\
	B &= b r^{-1},&  \quad p &= a r^{-(2/3) (\eta+2)} . \label{solns}
	\end{align}
	
A check for self-consistency shows that the density in the lab frame, given by $p^{3/4} \Gamma$, has no dependence upon $\theta$ and scales as $r^{-2}$, as is expected for nearly radial flow.

It should be noted that for the toroidal magnetic field and corresponding electric field to exist within the boundary layer, there must be a current distribution and charge distribution within the layer, and a current sheet and surface charge at the outer boundary where the magnetic and electric fields terminate. Calculating the current distribution within the jet from the solutions in Eq \eqref{solns}, one can see that longitudinal current within the jet is conserved in the case of approximately radial streamlines, providing another check of self-consistency.

% ===================================
\begin{figure}
\center
\includegraphics[width=3.0in]{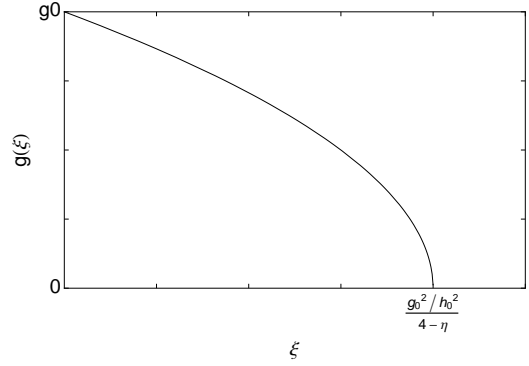}
\caption{Special-case solution for $g(\xi)$ in terms of the pressure-profile parameter $\eta$ and boundary conditions $g_0$ and $h_0$.}
\label{fig: gplot}
\end{figure}
% ===================================

%%%%%%%%%%%%%%%%%%%%%%%%%%%%%%
%%%%%%%%%%%%%%%%%%%%%%%%%%%%%%

\section{Discussion of Results}

The first important result of this solution is that, in the observer's frame, both the magnetic field $B$ and the transverse jet velocity  $\beta_\theta$ have only radial dependence; they are constant across the boundary layer. This is in direct opposition to the results from the strictly hydrodynamic limit (see KBB12, Section 3.1), where $\beta_\theta$ decreases monotonically from the outside of the boundary layer inward for all cases where $2<\eta<4$. 

Another significant point is that the solution for the magnetic field has no dependence in either dimension on the parameter $\eta$, meaning that the magnetic field that develops within the boundary layer is not affected by the pressure profile of the medium that the jet passes through.

The Lorentz factor $\Gamma$, on the other hand, does have a radial dependence on $\eta$. As expected, we see that the jet is accelerated as it propagates outward: $\Gamma \propto r^{\eta/2 - 1}$ scales as a positive power of $r$.

We now examine the pressure profile within the boundary layer as prescribed by this solution. Since we have demonstrated that the magnetically-dominated regime is the relevant regime in this problem, the total pressure within the layer is approximately described by the magnetic pressure
	\begin{align}
	\dfrac{B^2}{\Gamma^2} \propto b_0^2 h_0^2 \left[ \left(\dfrac{g_0}{h_0}\right)^2 - (4 - \eta) \xi \right]. \label{magpressure}
	\end{align}
 
In KBB12 we demonstrated that in the strictly hydrodynamic case, the pressure monotonically decreases for all $\eta$, but decreases linearly with $\xi$ only for the case where $\eta = 8/3$ (KBB12, Section 3.1). In this specific case, the pressure drops to zero within the boundary layer, indicating that all of the jet material is piled in a thin boundary layer in a `hollow cone' structure. As the value of $\eta$ gets further from $8/3$ in either direction, the pressure profile becomes less steep, implying that the boundary layer broadens and the structure of the jet becomes less like a hollow cone. 

The magnetic case is a little more difficult to interpret due to the unspecified boundary conditions $g_0$ and $h_0$ in Eq \eqref{magpressure}, but we can make some qualitative observations. First, it is clear that the magnetic case parallels the hydrodynamic case in that all solutions for pressure (which, since magnetically dominated, scales as $1/\Gamma^2$ in the boundary layer) monotonically decrease with $\xi$ for $2<\eta<4$. This means that the pressure is greatest at the contact discontinuity where it is matched with the external pressure, and it decreases inwards across the layer (see Fig. \ref{fig: gplot}), as is expected due to the collimation. 

A major contrast between the magnetic and hydrodynamic cases is that in the magnetic case it is true for  \emph{all} values of $\eta$ that the pressure decreases linearly and vanishes at a critical value of $\xi$, in this case $\xi_{cr} = \left(g_0 / h_0\right)^2/(4-\eta)$. Thus any jet can form a hollow cone structure in the magnetic case, with a boundary layer of thickness $\Delta \xi = \xi_{cr}$. The impact of $\eta$, then, is instead to determine the thickness of the boundary layer: jets injected into an ambient medium with a pressure profile parametrized by a low $\eta$ (a gradual decrease in external pressure) will therefore develop the most pronounced hollow-cone structure, whereas jets injected into a medium that has a higher $\eta$ (a steep decrease in external pressure) will exhibit a thicker boundary layer. 

The distinction between the magnetic and hydrodynamic cases is most evident, however, when examining the global energy constraints for these solutions, as in KBB12. In this case, we assume that all the power in the boundary layer is in the form of Poynting flux, given by $L \propto B^2  A$, where $A$ is the cross-sectional area of the boundary layer. As we are assuming a boundary layer of width $\Delta \theta = 1/\Gamma$, the cross-sectional area of that ring is $A \propto r^2 / \Gamma$, implying that the power within the boundary layer scales as $L \propto r^{1 - \eta/2}$. Thus for  $2<\eta<4$, the boundary layer contains less and less energy as one goes further out in $r$. 

This loss of energy is indicative of a unique situation in the magnetic case: rather than being bounded on the inside by a shock front through which material is added, as in the hydrodynamic case, at large $r$ the boundary layer instead appears to be bounded on the inside by a rarefaction front, through which material is leaving the layer and returning to the main jet.

We can verify the position of this rarefaction front by calculating the location of the surface at which the fluid motion normal to the surface is equivalent to the local speed of sound. This condition can be stated as
	\begin{align}
	\beta_{\xi_r}^2 = \dfrac{\beta_f^2}{\Gamma^2 \left( 1 - \beta_f^2 \right)},
	\end{align}
where $\beta_{\xi_r}$ is the velocity normal to the surface of the rarefaction front $\xi_r$, and $\beta_f$ is the local sound speed, which is given by the relativistic expression for the fast magnetosonic speed in the hot-jet limit,
	\begin{align}
	\beta_f^2 = \dfrac{\frac{4}{3} p + B^2}{4 p + B^2}.
	\end{align}

This calculation yields an approximate position for the rarefaction front of
	\begin{align}
		\xi_r = \xi_{cr} - \dfrac{(\eta-2)^{3/2}}{(4-\eta)^{5/2}} \left(\dfrac{2 a_0}{3 b_0^2}\right)^{3/4} \left( \dfrac{g_0}{h_0} \right)^{1/2}  r^ {(\eta-4)/4},
	\end{align}
indicating that the rarefaction front occurs at a location just before the pressure within the boundary layer drops to zero, allowing matching across the front to the conditions in the interior of the jet. At large $r$, the position of the rarefaction front asymptotes to the location of the pressure zero-point.

Looking at how the Lorentz factor $\Gamma$ scales with $r$ along the rarefaction front, we see that $\Gamma \propto r^{3 \eta / 8 - 1/2}$. We know that the Lorentz factor scales as $\Gamma \propto r^{\eta/2 - 1}$ at the contact discontinuity, and we expect it to scale as $\Gamma \propto r$ in the jet interior  to be consistent with free expansion (see also \citealp{Lyubarsky11}). Thus, for $2 < \eta < 4$, the scaling of $\Gamma$ at the rarefaction front is consistent with an intermediate acceleration between the two, providing a smooth transition between the boundary layer and the jet interior.

More insight into the boundary-layer energy loss can be attained by examining the position of the contact discontinuity ($\theta_c$) and the inner boundary ($\theta_s = \theta(g=0)$) as a function of radius. Using the solutions found in Eq \eqref{solns}, the width of the boundary layer is given by
	\begin{align}
	\theta_c - \theta_s = - \dfrac{g_0^2/h_0}{4-\eta}r^{1-\eta/2},
	\end{align}
indicating that the width is decreasing in $r$.

Furthermore, integrating Eq \eqref{dthetac} using the above equation allows us to find the form of the inner and outer boundaries individually:
	\begin{align}
	\theta_c &= \theta_0 - \dfrac{h_0}{\eta/2 - 1} r^{-(\eta/2 - 1)}\\
	\theta_s &= \theta_0 + \left( \dfrac{h_0}{4-\eta} \right)\left[ \left(\dfrac{g_0}{h_0}\right)^2 - \dfrac{4-\eta}{\eta/2 - 1} \right]  r^{-(\eta/2 - 1)} .
	\end{align}
 From this, we can see that the outer boundary is collimating, but less and less so with increasing $r$. The inner boundary, on the other hand, is \em decollimating \em when the condition 
	\begin{align}
	\left(\dfrac{g_0}{h_0}\right)^2 > \dfrac{4-\eta}{\eta/2 - 1}
	\end{align}
is met, with the boundary conditions prescribed by matching across the rarefaction front to solutions for the jet interior (such as those described by \citealp{Lyubarsky11}). When this condition is satisfied, the boundary layer intercepts fewer and fewer streamlines of new material from within the jet interior. The decollimation, too, weakens with increasing $r$, with the inner and outer boundaries meeting asymptotically.

These results suggest that the boundary layer of a magnetically dominated jet decreases in width with increasing distance from the source, and contains a decreasing amount of energy as material leaves the boundary layer across the rarefaction front to rejoin the jet interior. Nonetheless, a sharp pressure gradient is maintained across the layer, insulating the interior of the jet from the external medium.

%%%%%%%%%%%%%%%%%%%%%%%%%%%%%%
%%%%%%%%%%%%%%%%%%%%%%%%%%%%%%

\section{Conclusion}

We have evaluated the structure of a boundary layer within a hot, magnetohydrodynamic jet with an ultrarelativistic equation of state. We assumed that the jet as a whole is causally disconnected from its surroundings, but the boundary layer is thin and therefore in causal contact. We examined the impact on jet collimation of a toroidal magnetic field within the jet as well as an ambient medium with a pressure profile of $p \propto r^{-\eta}$, with $2<\eta<4$.

We first demonstrated that the basic RMHD equations can be used to show that any jet boundary layer seeded with a toroidal magnetic field at its base will eventually become magnetically dominated at large radii.

We then constructed self-similar solutions for the boundary layer in the limit where the jet pressure is dominated by magnetic pressure. We found a special case of physical solutions where the jet pressure decreases linearly across the boundary layer, dropping to zero at a location set by the boundary conditions and the value of the pressure profile parameter $\eta$. The boundary layer thickness is dependent upon the value of $\eta$, with increasing $\eta$ producing an increasingly wide boundary layer.

We further found that the thickness of the boundary layer decreases with radial distance, and the boundary layer contains a decreasing amount of the jet energy. This suggests that the addition of a magnetic field fundamentally changes the jet at large radius: whereas in the hydrodynamic case the inner boundary of the layer is a shock front through which material enters the layer, in the magnetohydrodynamic case the layer is bounded on the inside by a rarefaction front through which material leaves the boundary layer and rejoins the interior of the jet.

We found the position of this rarefaction front to occur just inside the layer from the location where the pressure vanishes, providing a smooth transition between the boundary layer and the jet interior and allowing for matching across the rarefaction front to the conditions in the jet interior. This matching would prescribe the values of the boundary conditions within the layer, and could potentially yield a solution where the rarefaction front is gradually decollimating, intercepting fewer and fewer streamlines as radial distance from the source increases.

In spite of the thinning of the boundary layer with radial distance, a sharp pressure gradient is nonetheless maintained across the layer, causing it to function as an insulating buffer between the jet interior and the ambient medium. Unlike the hydrodynamic case, the solutions for the structure of a magnetized jet do not have clear observational implications. Though the boundary layer contains a decreasing amount of energy as one looks further from the source, the layer might nonetheless have a high emissivity, which could be observationally important if the flow within the boundary layer is pointed along our line of sight.

The results presented in this paper provide the premise for a more complete treatment in numerical simulations of the effects of the ambient medium on collimation, both by demonstrating the behavior of the jet when the outer wall is allowed to change its shape, and by providing models that can be used to assess the effects of numerical resolution on simulation outcomes.

Ultimately, these results provide a foundation for future work examining energy dissipation in magnetized jets and the associated radiative observational signatures.

%%%%%%%%%%%%%%%%%%%%%%%%%%%%
%%%%%%%%%%%%%%%%%%%%%%%%%%%%
%%%%%%%%%%%%%%%%%%%%%%%%%%%%

\section*{Acknowledgements}

This work was supported in part by NSF grant AST-0907872, NASA Astrophysics Theory Program grant NNX09AG02G, and NASA's Fermi Gamma-ray Space Telescope Guest Investigator program.
 \\

%%%%%%%%%%%%%%%%%%%%%%%%%%%
%%%%%%%%%%%%%%%%%%%%%%%%%%%

%\bibliographystyle{hapj}
%\bibliography{refs}

%%%%%%%%%%%%%%%%%%%%%%%%%%%%
%%%%%%%%%%%%%%%%%%%%%%%%%%%%

\label{lastpage}

\end{document}